\documentclass{mem_l}
\usepackage{natbib}\usepackage{txfonts}\usepackage{balance}
\usepackage{graphicx}
\usepackage[a4paper]{hyperref}
\idline{75}{282}
\begin{document}
\def\teff{$T\rm_{eff }$}
\def\kms{$\mathrm {km s}^{-1}$}
\def\sgr{Sgr dSph}
\def\asf{[$\alpha$/Fe]}
\def\fsh{[Fe/H]}

%%%%%%%%%%%%%%%%%%%%%%%%

%%%%%%%%%%%%%%%%%%%%%%%%

\title{
The chemistry of the neighbors: detailed abundances in the Sgr and CMa dwarf galaxies
}

   \subtitle{}

\author{
L. \,Sbordone\inst{1,3}
P. \,Bonifacio\inst{1}
G. \,Marconi\inst{2}
S. \,Zaggia\inst{1}
\and R. \,Buonanno\inst{4} 
          }

  \offprints{L. Sbordone}

\institute{
INAF --
Osservatorio Astronomico di Trieste, Italy
\and
ESO -- European Southern Observatory, Santiago, Chile
\and
INAF --
Osservatorio Astronomico di Roma,  Italy 
\and
Universit\`a di Roma 2 ``Tor Vergata'', Rome, Italy
\\
\email{sbordone@oa-roma.inaf.it}
}

\authorrunning{Sbordone et al.}

\titlerunning{Abundances in Sgr and CMa}

\abstract{
We summarize the results of our ongoing investigation of the chemical abundances in the Sagittarius Dwarf Spheroidal Galaxy (\sgr) and in the Canis Major Overdensity (CMa). 12 RGB stars were analyzed in the Sgr dSph, plus 5 in the associated globular Terzan 7, together with three CMa candidate members. Detailed abundances have been derived for up to 23 elements from Oxygen to Europium. 

}
\maketitle{}

\section{Introduction}

In recent years, Local Group dwarf galaxies have acquired an increasing importance since they are believed to constitute the relic ``building blocks'' out of which major galaxies should have formed. In particular, the discovery of \sgr\ \citep{ibata95}, and the subsequent detection of its tidal debris in the Halo \citep[see e.g.][]{majewski03} clearly showed how dwarf galaxies still play an important role in the evolution of major galaxies like the Milky Way. More recently, the tidal structure known as GASS \citep[Galactic Anticenter Stellar Structure, see][]{newberg02,crane03} has been associated to the still controversial detection of a dwarf galaxy remnant in CMa \citep[see][]{martin04,bellazzini04,momany04}.

\section{Sgr dSph and Terzan 7}
We are conducting a study on the chemical abundances in \sgr\ and in the associated globular cluster Terzan 7, by analyzing high resolution VLT-UVES spectra of RGB stars, and deriving abundances for up to 23 elements from Oxygen to Europium. So far, results for Iron and $\alpha$ elements have been published in \citet{bonifacio04} for 12 \sgr\ main body stars (T$_{eff} \sim$ 4900 K and $\log g \sim$ 2.5). For two of them an analysis for 21 elements has been presented in \citet{bonifacio00}. In \citet{sbordone05ter7} we analyze 5 Terzan 7 stars (T$_{eff}$ between 3945 and 4421 K, $\log g$ between 0.8 and 1.3). Sulfur abundances for three of the Ter 7 stars have been presented in \citet{caffau05}. In all cases \citep[except in][]{bonifacio00}, we have employed our Linux-ported version of the ATLAS-WIDTH-SYNTHE suite \citep{kurucz93,sbordone04atlas,sbordone05atlas} to compute stellar atmosphere models and spectral syntheses, and to derive abundances. 

The studied stars in the \sgr\ main body show high metallicity (\fsh\ between -0.8 and solar) and an undersolar \asf\ ratio, decreasing with increasing metallicity. \sgr\ appears to fall on the same \fsh\ vs \asf\ sequence with the other LG dSph, populating the Fe-rich, $\alpha$-poor end of the sequence. This indicates that a prolonged albeit slow star formation has taken place in the galaxy, a fact confirmed also by the very young age of the studied population (age $<$ 2 GYr). We looked for traces of interstellar gas in \sgr, to support for such a recent star formation, finding only little traces of ISM \citep[see][]{monai05}. Also the other studied elements present many significant departures from solar ratios, with underabundant Na, Al and Sc, deficient Ni, Cu and Zn, and strongly overabundant La, Ce and Nd. 
The five Ter 7 stars appear to share the same ``chemical signature'', showing a mean \fsh\=-0.59 and solar \asf. 

The chemical pattern displayed by the young \sgr\ population can be used to probe stellar populations, now belonging to the Milky Way (MW), to verify if they formed inside the \sgr\ system. A striking example of this is Palomar 12, a young MW globular cluster, which abundances have been measured by \citet{cohen04}. This cluster was already suspected to have been tidally stripped from \sgr\ \citep[e. g.][]{bellazzini03}. Actually, Pal 12 chemical pattern appears to strikingly reproduce the one we find in \sgr, even in the most significant departures from the solar ratios (such as $\alpha$ elements, Al, Ni, Cu, Zn, La, Ce...), thus leading to consider essentially sure its origin inside the \sgr\ system.

\section{The CMa overdensity}
Shortly after the detection of the CMa overdensity, we undertook an exploratory program based on VLT-FLAMES spectra, to study the chemical abundances of CMa candidate member stars. In \citet{sbordone05cma} we present the results on the three UVES stars that appeared suitable as CMa members, analyzed with the same techniques described above. Based on chemical abundances, one of them is most likely a MW interloper, a second one is of uncertain origin, while the third one shows a rather peculiar abundance pattern that makes it unlikely to have originated inside the MW. It is a metal rich (\fsh=0.15) subgiant (T$_{eff}$=5367 K, $\log g$=3.5), showing again underabundant $\alpha$ elements, high La, Ce and Nd abundance, but, at odds with what we find in \sgr, a significant Cu overabundance ([Cu/Fe]=0.25). The radial velocity of this star, also, appears to be the one that satisfies better the reported dynamics of the CMa overdensity.
%

% \begin{acknowledgements}
% I am grateful to R. Gratton and to all the researchers
% of the Large Programme led by him, for the excellent 
% work done and for allowing me to present some
% results in advance of publication.
% It is a pleasure to  acknowledge the
% many helpful discussions on the topic
% of Li abundances with M. Spite, F. Spite, P. Molaro and
% R. Cayrel.
% \end{acknowledgements}

\bibliographystyle{aa}

\end{document}